\documentclass[12pt]{article}
\usepackage{lscape}
\usepackage{cite}
\usepackage{graphicx,graphics}
\usepackage{amssymb}
\usepackage{amsmath}
\usepackage{dcolumn}
\usepackage{bm}
\usepackage{color}
\def\be{\begin{equation}}
\def\ee{\end{equation}}
\newcommand{\bea}{\begin{eqnarray}}
\newcommand{\eea}{\end{eqnarray}}
\newcommand{\nn}{\nonumber}
\newcommand{\eV}{\mbox{eV}}

\newcommand{\GeV}{\mbox{GeV}}

\def\hbar#1{\backslash\hspace{-2mm}#1}

\def\nn{\nonumber}

\def\2tvec#1#2{
\left(
\begin{array}{c}
#1  \\
#2  \\
\end{array}
\right)}
\def\mat2#1#2#3#4{
\left(
\begin{array}{cc}
#1 & #2 \\
#3 & #4 \\
\end{array}
\right) }
\def\Mat3#1#2#3#4#5#6#7#8#9{
\left(
\begin{array}{ccc}
#1 & #2 & #3 \\
#4 & #5 & #6 \\
#7 & #8 & #9 \\
\end{array}
\right) }
\def\3tvec#1#2#3{
\left(
\begin{array}{c}
#1  \\
#2  \\
#3  \\
\end{array}
\right)}

\def\hbar#1{\backslash\hspace{-2mm}#1}

\def\nn{\nonumber}

\newcommand{\bt}{\begin{itemize}}
\newcommand{\et}{\end{itemize}}
\def\lsim{\raise0.3ex\hbox{$\;<$\kern-0.75em\raise-1.1ex\hbox{$\sim\;$}}}
\def\gsim{\raise0.3ex\hbox{$\;>$\kern-0.75em\raise-1.1ex\hbox{$\sim\;$}}}
\numberwithin{equation}{section}

\begin{document}

\begin{titlepage}

\begin{center}

\vspace{1cm}
{\large\bf {One-parameter Neutrino Mass Matrix and Symmetry Realization}}
\vspace{2cm}

Yuji Kajiyama{\footnote{kajiyama-yuuji@akita-pref.ed.jp}},
Akari Sato, Wakako Sato and Aya Suzuki
\vspace{5mm}

{\it%
Akita Highschool, Tegata-Nakadai 1, Akita, 010-0851, Japan
}
  
  \vspace{8mm}

\abstract{
We investigate the Majorana neutrino mass matrix $M_{\nu}$ with one parameter  in 
the context of two texture zeros and its symmetry realization by 
non-Abelian discrete symmetry. From numerical calculation, 
we confirm that the textures $(M_{\nu})_{11,12}=0$ and $(M_{\nu})_{11,13}=0$ 
are consistent with the current experimental constraints, and show the correlations 
between non-zero elements of $M_{\nu}$. 
The ratios of non-zero elements of $M_{\nu}$ are constrain in small regions, 
and we find simple examples of $M_{\nu}$ with one real mass parameter. 
We also discuss symmetry realization of the mass matrix by 
the type-II seesaw mechanism based on the binary icosahedral symmetry $A_5'$ .
 }

\end{center}
\end{titlepage}

\setcounter{footnote}{0}

\section{Introduction}
The recent precise measurements for the neutrino sector indicate that the 
mixing angle $\theta_{13}$ has finite non-zero value
\cite{Abe:2011sj,Adamson:2011qu,Abe:2011fz,An:2012eh,Ahn:2012nd}.  
This fact indicates the modification of models such as Tri-Bi Maximal mixing
 \cite{tribimaximal} which derives $\theta_{13}=0$. Several ideas to overcome this 
 problem have been proposed so far, based on flavor symmetries \cite{Altarelli:2010gt,Ishimori:2010au}, perturbation of Pontecorvo-Maki-Nakagawa-Sakata (PMNS)  matrix $U_{PMNS}$ from symmetric textures 
 \cite{Wang:2013wya}, texture zeros of 
 neutrino mass matrix $M_{\nu}$
  \cite{Frampton:2002yf,Meloni:2012sx, Grimus:2012zm, 
  Fritzsch:2011qv,Lashin:2011dn}, anarchy \cite{Hall:1999sn, Gluza:2011nm} and 
  vanishing minors of  $M_{\nu}$ \cite{vanishingminor, Araki:2012ip} etc.

For models of two texture zeros in $M_{\nu}$, with four independent parameters, 
it has been shown 
\cite{Meloni:2012sx} that 
the following two patterns 
\be
{\rm A}_1~:~M_{\nu}({\rm A}_1)=\left(\begin{array}{ccc}
0&0&\times\\
0&\times &\times\\
\times&\times&\times\\
\end{array}\right),~~
{\rm A}_2~:~M_{\nu}({\rm A}_2)=\left(\begin{array}{ccc}
0&\times&0\\
\times&\times &\times\\
0&\times&\times\\
\end{array}\right),
\label{a1a2}
\ee 
where $\times$ denotes non-zero matrix elements, 
require less fine-tuning of parameters to satisfy the current experimental bounds. Moreover it is discussed in Ref. \cite{Grimus:2012zm} that 
if all parameters are real and if ratio of non-zero 
matrix elements are small integer (1 or 2), textures 
given in Eq. (\ref{a1a2}) with two independent parameters 
are in good agreement with the experimental data. 

In this paper, we consider a mass matrix $M_{\nu}$ with {\it one} real parameter 
in the case of textures Eq. (\ref{a1a2}), such as 
\be
M_{\nu}=m\left( \begin{array}{ccc}
0&0&1\\
0&2&3\\
1&3&3\\
 \end{array}\right), 
 \label{mnu0}
\ee  
where $m$ is a mass parameter determined by experimental constraints of 
two mass-squared differences $\Delta m_{21}^2$ and $\Delta m_{31}^2$ of 
neutrinos. First we perform numerical calculation in the case of two texture zeros 
A$_1$ and A$_2$ with four real parameters, and find 
several candidates of $M_{\nu}$ with one parameter by assuming that 
ratio of matrix elements are small real number (including integer). After that, 
we discuss symmetry realization of a candidate Eq.(\ref{mnu0}) by 
non-Abelian discrete symmetry $A_5'$\cite{Hashimoto:2011tn}. 
The $A_5'$ symmetry contains three- and five-dimensional irreducible representation 
${\bf 3}$ and ${\bf 5}$, and the singlet ${\bf 1}$ from their tensor product 
${\bf 3 \cdot 5 \cdot 3}={\bf 1}+\cdots$ enters all the elements of $M_{\nu}$ with 
desired weights if one assigns ${\bf 3}$ and ${\bf 5}$ for neutrinos and 
Higgs bosons, respectively
{ \footnote{ In Ref.\cite{Everett:2008et}, the $A_5$ symmetry 
is discussed which has similar feature.} }. 

This paper is organized as follows.
In the next section, we perform numerical calculation of 
neutrino mass matrix $M_{\nu}$ given in Eq.(\ref{a1a2}), and show allowed regions 
of non-zero matrix elements. 
Several explicit forms of $M_{\nu}$ with one real free parameter 
are also given. In Section 3, we discuss symmetry realization of a concrete example 
of $M_{\nu}$ given in Eq.(\ref{mnu0}) by the binary icosahedral symmetry $A_5'$. 
Section 4 is devoted to the conclusions. In Appendix, the Higgs potential 
in our model based on the $A_5'$ symmetry is given. 
\section{One-parameter Texture of $M_{\nu}$}
In this section, we first perform numerical calculation for Majorana 
neutrino mass matrix $M_{\nu}$ with two zero entries in order to 
find textures with one real free parameter.  
Next, we give some examples of $M_{\nu}$ with one parameter based on the
numerical calculation and related quantities such as 
the PMNS matrix and mass squared differences. 
From a standpoint of flavor symmetry, it is preferable that ratios of 
non-zero elements of $M_{\nu}$ are simple small real number.

\subsection{Numerical Calculation in Two Texture Zeros}
We assume two zero entries in $M_{\nu}$ \cite{Frampton:2002yf} in our numerical calculation. 
As discussed in Refs.\cite{Meloni:2012sx,Grimus:2012zm,Fritzsch:2011qv}, if one includes the CP-violating phases, the textures given in Eq.(\ref{a1a2})
require less fine-tuning of parameters to satisfy the current experimental constraints given below, 
although seven patterns of $M_{\nu}$, such as A$_{1,2}$, B$_{1,2,3,4}$ and 
C,{\footnote{Since we focus on the 
pattern A$_{1}$ and A$_{2}$ due to the conclusions of Refs. 
\cite{Meloni:2012sx,Grimus:2012zm}, we do not show the explicit form of 
patterns B$_{1,2,3,4}$ and C. See Refs. \cite{Frampton:2002yf,Fritzsch:2011qv} for 
the definition of those.}} are consistent with the experiments
\cite{Fritzsch:2011qv} . 
Moreover if all parameters are real, the patterns A$_1$ and A$_2$ with 
two free parameters show good agreement with the experiments \cite{Grimus:2012zm}.  
Two matrices in Eq.(\ref{a1a2}) are related to each other by the relation
\bea
PM_{\nu}({\rm A}_1)P=M_{\nu}({\rm A}_2),~~{\rm with}~~
P=\left(\begin{array}{ccc}
1&0&0\\
0&0 &1\\
0&1&0\\
\end{array}\right), 
\eea 
and the both cases lead to the Normal Hierarchy (NH) of the neutrino masses 
$(m_3>m_2>m_1)$.

In order to find texture of $M_{\nu}$ with one real free parameter, we perform numerical calculation in the following procedure; 

\begin{enumerate}
\item We assume all parameters in $M_{\nu}$ to be real, and choose the basis 
in which the left-handed charged leptons are mass eigenstates. 
The PMNS matrix $U_{PMNS}$ 
with vanishing CP-violating phases and 
neutrino mass eigenvalues $(m_1,m_2,m_3)$ are defined as 
\bea
M_{\nu}=U_{PMNS}\left( \begin{array}{ccc}
m_1&&\\&m_2&\\&&m_3\\
\end{array}\right)U_{PMNS}^T,
\label{diag}
\eea
where $m_2=\pm \sqrt{\Delta m_{21}^2+m_1^2}$ and 
$m_3=\pm \sqrt{\Delta m_{31}^2+m_1^2}$ with $\Delta m_{ij}^2=m_i^2-m_j^2$. 
\item We focus on the pattern A$_1$ and A$_2$ given in Eq.(\ref{a1a2}). 
\item The global fit data \cite{GonzalezGarcia:2012sz} for the 
case of NH at the 3$\sigma$ level \footnote{See also Refs. \cite{Tortola:2012te,Fogli:2012ua} for the other global fit results.} ,
\bea
&&\sin^2 \theta_{12}=[0.267,0.344],~
\sin^2 \theta_{23}=[0.342,0.667],~
\sin^2 \theta_{13}=[0.0156,0.0299],\nn\\&&
\Delta m_{21}^2=[7.00,8.09] \times 10^{-5}~\eV^2,~
\Delta m_{31}^2=[2.276,2.695] \times 10^{-3}~\eV^2,
\label{exp}
\eea
are used. We randomly input the above constraints except $\Delta m_{31}^2$ into the right-hand side of Eq.(\ref{diag}), and find $m_1$ and $\Delta m_{31}^2$ 
by solving equations $(M_{\nu})_{11}=0$ and $(M_{\nu})_{12(13)}=0$ for 
the case A$_{1(2)}$.
 \end{enumerate} 
\begin{figure}[h]
\begin{center}
 \includegraphics[width=87mm]{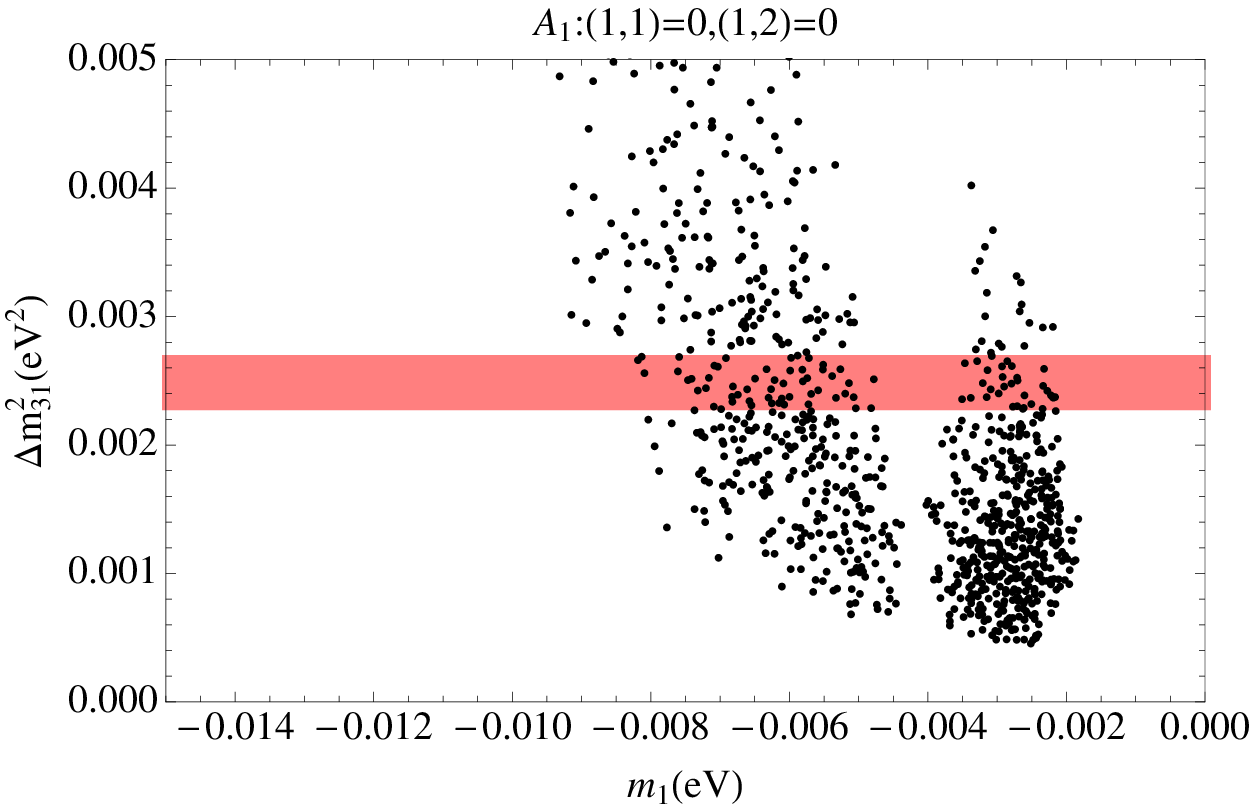}
 \includegraphics[width=87mm]{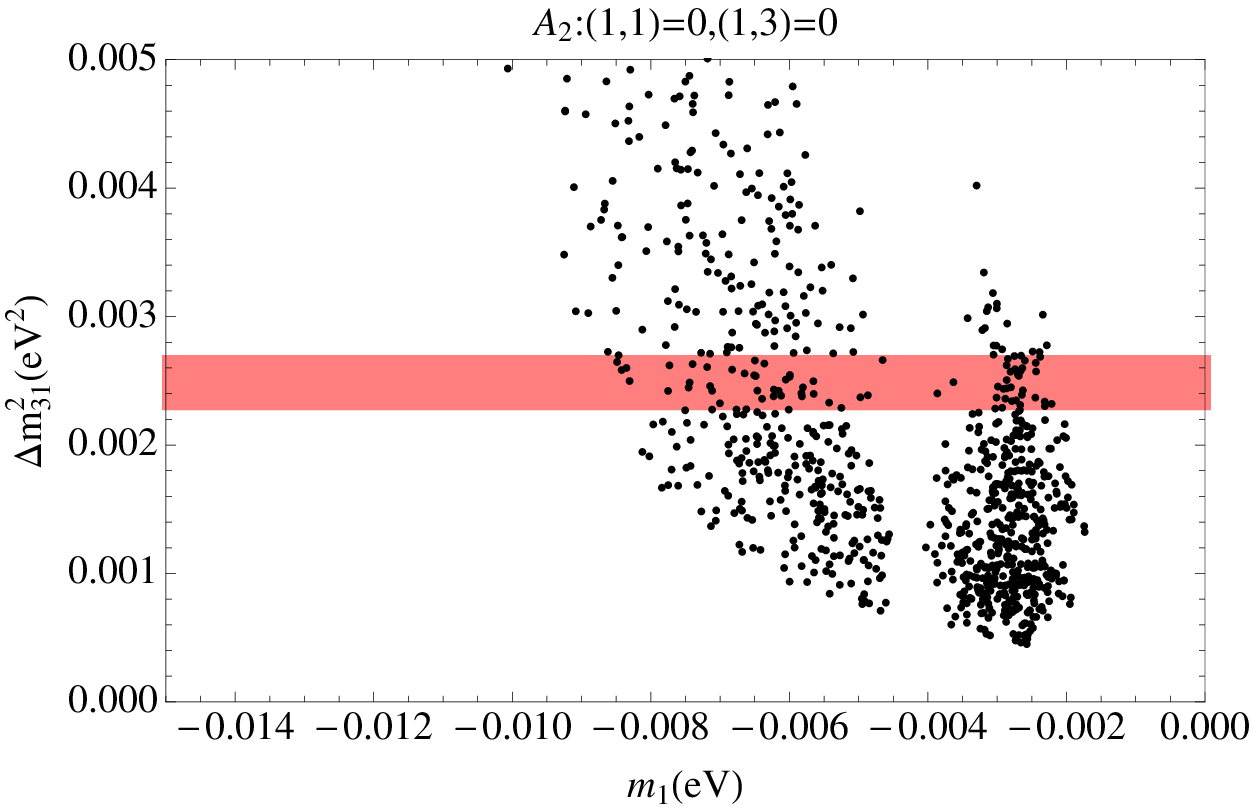}
   \vspace{-1cm}
 \caption{Neutrino mass $m_1$ and $\Delta m_{31}^2$ in the case of 
 $(M_{\nu})_{11}=(M_{\nu})_{12}=0$ (left panel) and 
  $(M_{\nu})_{11}=(M_{\nu})_{13}=0$ (right panel). In both panels, 
  allowed region of $\Delta m_{31}^2$ given in Eq.(\ref{exp}) is represented by light-red (dark) regions.}
   \label{fig1}
\end{center}
\end{figure}

Figure \ref{fig1} shows the results for the case of 
A$_1~:~(M_{\nu})_{11}=(M_{\nu})_{12}=0$ (left panel) and 
A$_2~:~(M_{\nu})_{11}=(M_{\nu})_{13}=0$ (right panel) in 
the $m_1-\Delta m_{31}^2$ plane.
In both panels, allowed region of $\Delta m_{31}^2$ given in Eq.(\ref{exp})
is represented by the light-red (dark) regions. All the dots in both panels satisfy the global fit constraints 
Eq.(\ref{exp}) except $\Delta m_{31}^2$, and one can see that there exist dots in the allowed region of $\Delta m_{31}^2$. 
Therefore we confirm that both the cases A$_{1}$ and A$_{2}$ are 
consistent with the current experimental bounds\cite{Meloni:2012sx,Grimus:2012zm}.

\begin{figure}[h]
\begin{center}
 \includegraphics[width=87mm]{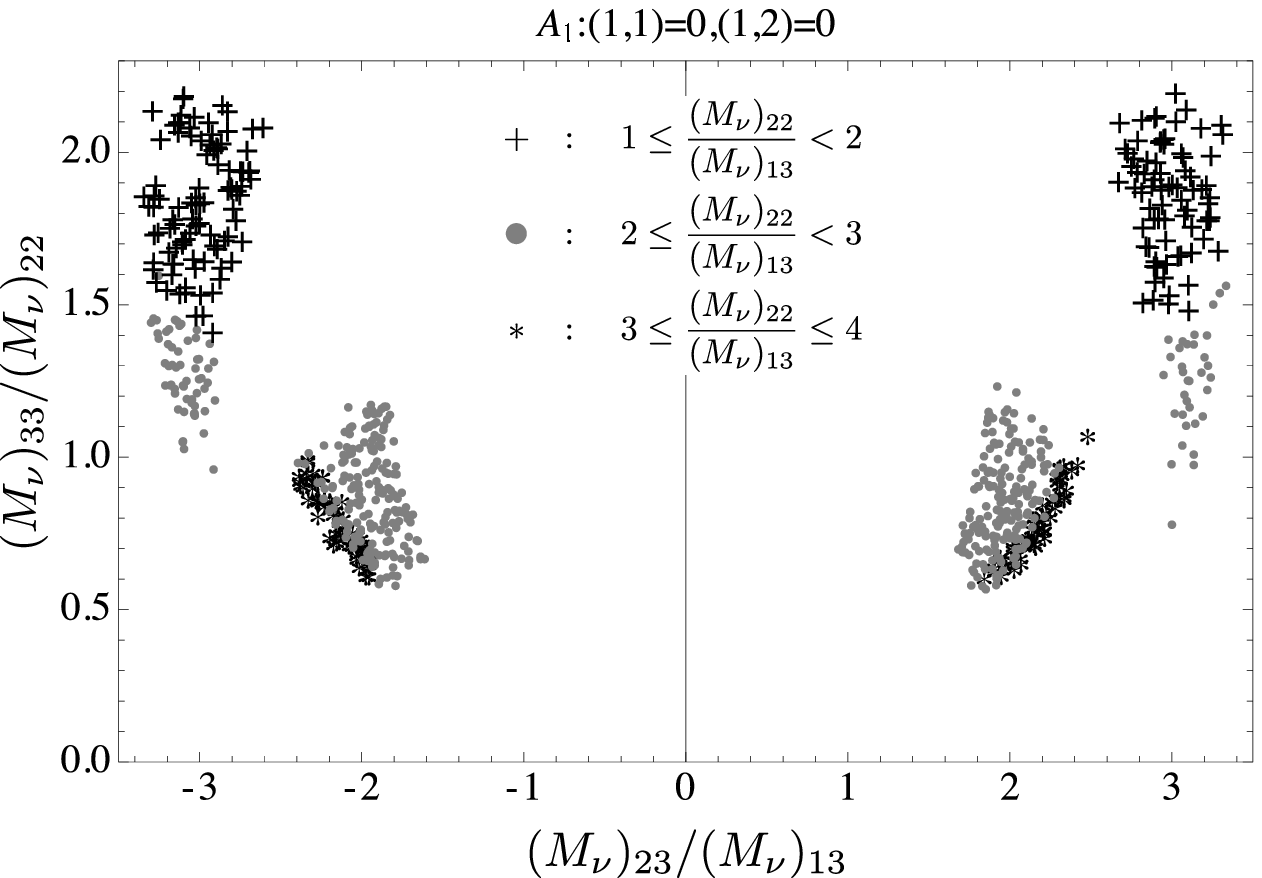}
\hspace{3mm}
 \includegraphics[width=87mm]{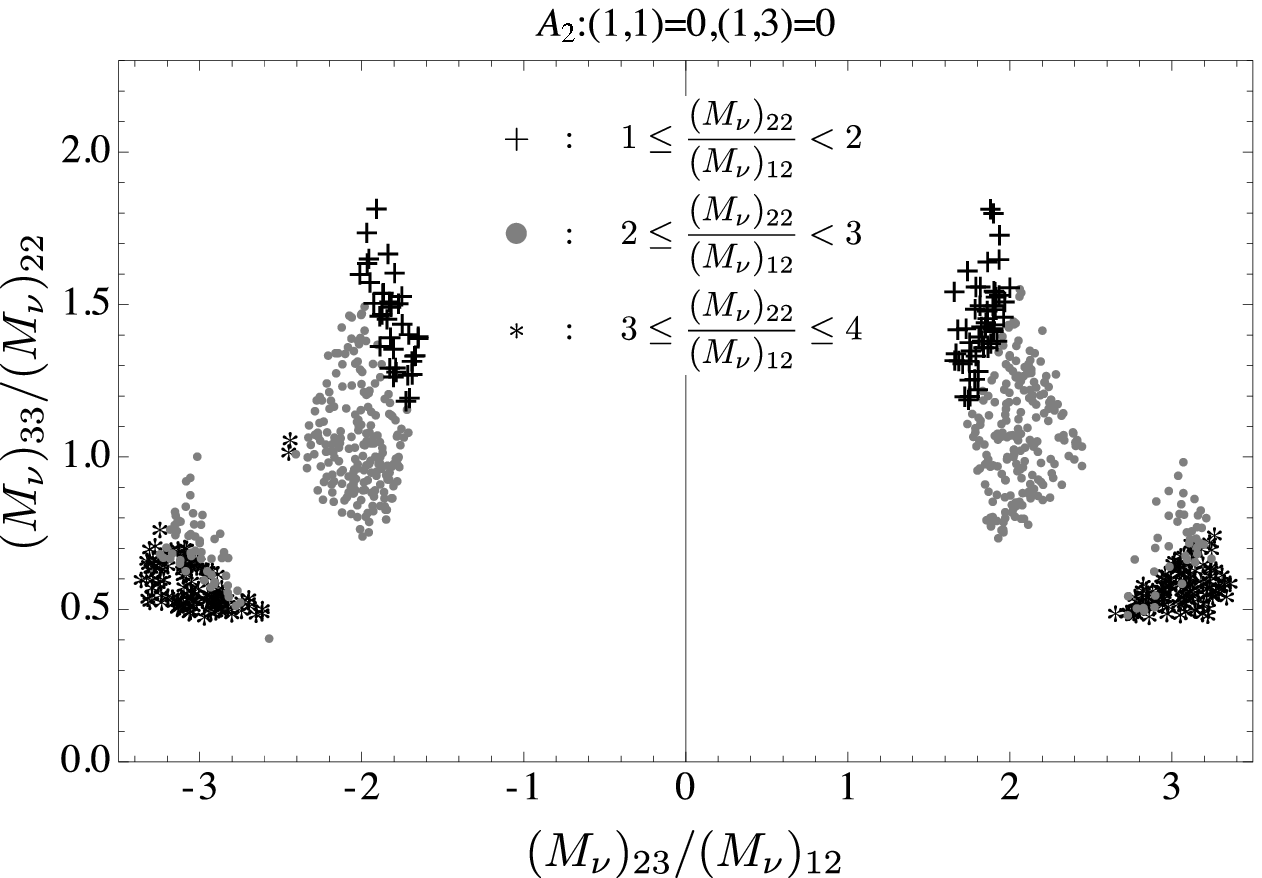}
   \vspace{-1cm}
 \caption{Correlations between $(M_{\nu})_{23}/(M_{\nu})_{13}$ 
and $(M_{\nu})_{33}/(M_{\nu})_{22}$ (left panel) for the case of A$_1$, 
and $(M_{\nu})_{23}/(M_{\nu})_{12}$ 
and $(M_{\nu})_{33}/(M_{\nu})_{22}$ (right panel) for the case of A$_2$. 
The symbols $+$, grey points and $\ast$ represent $1 \leq |(M_{\nu})_{22}/ (M_{\nu})_{13(12)}|<2$, $2 \leq |(M_{\nu})_{22}/ (M_{\nu})_{13(12)}|<3$ and 
$3 \leq |(M_{\nu})_{22}/ (M_{\nu})_{13(12)}|\leq 4$ for the left (right) panel, respectively.}
   \label{fig2}
\end{center}
\end{figure}

Next we discuss correlations between non-zero elements of $M_{\nu}$. 
Figure \ref{fig2} shows the allowed region in $(M_{\nu})_{23}/(M_{\nu})_{13}-
(M_{\nu})_{33}/(M_{\nu})_{22}$ plane (left panel) for the case of A$_1$, 
and in $(M_{\nu})_{23}/(M_{\nu})_{12}-(M_{\nu})_{33}/(M_{\nu})_{22}$ plane 
(right panel) for the case of A$_2$. 
In both panels, we have chosen the non-zero elements such that all experimental constraints Eq. (\ref{exp}) are satisfied. 
The symbols $+$, grey points and $\ast$ represent 
 $1 \leq |(M_{\nu})_{22}/ (M_{\nu})_{13(12)}|<2$, 
$2 \leq |(M_{\nu})_{22}/ (M_{\nu})_{13(12)}|<3$ and 
$3 \leq |(M_{\nu})_{22}/ (M_{\nu})_{13(12)}|\leq4$, respectively for 
the case A$_1$(A$_2$). There are no solutions for 
$|(M_{\nu})_{22}/ (M_{\nu})_{13(12)}|<1$ and 
$4<|(M_{\nu})_{22}/ (M_{\nu})_{13(12)}|$. 
From these figures, one finds solutions 
at $|(M_{\nu})_{23}/(M_{\nu})_{13(12)}|\sim 2~{\rm or}~3$, and  
$0.5 \lsim(M_{\nu})_{33}/(M_{\nu})_{22}\lsim2$ with 
$1<|(M_{\nu})_{22}/ (M_{\nu})_{13(12)}|<4$.



\begin{table}[h]
\centering {\fontsize{9}{12}
\begin{tabular}{|c||c|c|c|c|}
\hline  $M_{\nu}$ & 
$m\left(\begin{array}{ccc} 
0&0&1\\
0&2&3\\
1&3&3\\
\end{array}\right)$ &
 $m\left(\begin{array}{ccc} 
0&0&1\\
0&\sqrt{3}&2\sqrt{2}\\
1&2\sqrt{2}&2\sqrt{2}\\
\end{array}\right)$ &
 $m\left(\begin{array}{ccc} 
0&0&1\\
0&\sqrt{3}&2\sqrt{2}\\
1&2\sqrt{2}&3\\
\end{array}\right)$ &
 $m\left(\begin{array}{ccc} 
0&0&1\\
0&2&3\\
1&3&2\sqrt{2}\\
\end{array}\right)$  \\\hline
$U_{PMNS}$ &  
\!\!$\left(\begin{array}{ccc} 
\!\!\!\!\frac{3+\sqrt{7}}{2\sqrt{7+2\sqrt{7}}}&\!\!\!\!\frac{1}{\sqrt{3}}&\!\!\!\!\frac{3-\sqrt{7}}{2\sqrt{7-2\sqrt{7}}}\\
\!\!\!\!-\frac{1+\sqrt{7}}{2\sqrt{7+2\sqrt{7}}}&\!\!\!\!\frac{1}{\sqrt{3}}&\!\!\!\!-\frac{1-\sqrt{7}}{2\sqrt{7-2\sqrt{7}}}\\
\!\!\!\!\frac{1}{\sqrt{7+2\sqrt{7}}}&\!\!\!\!-\frac{1}{\sqrt{3}}&\!\!\!\!\frac{1}{\sqrt{7-2\sqrt{7}}}
\!\!\!\!\\
\end{array}\right)\!\!$ &
\!\!$\left(\begin{array}{ccc} 
\!\!\!\!0.815&\!\!\!\!0.560&\!\!\!\!0.147\\
\!\!\!\!-0.518&\!\!\!\!0.592&\!\!\!\!0.617\\
\!\!\!\!0.259&\!\!\!\!-0.579&\!\!\!\!0.773\!\!\!\\
\end{array}\right)$\!\!& 
\!\!$\left(\begin{array}{ccc} 
\!\!\!\!0.803&\!\!\!\!0.578&\!\!\!\!0.145\\
\!\!\!\!-0.534&\!\!\!\!0.589&\!\!\!\!0.606\\
\!\!\!\!0.265&\!\!\!\!-0.565&\!\!\!\!0.782\!\!\!\\
\end{array}\right)$\!\!  & 
\!\!$\left(\begin{array}{ccc} 
\!\!\!\!0.819&\!\!\!\!0.559&\!\!\!\!0.136\\
\!\!\!\!-0.504&\!\!\!\!0.581&\!\!\!\!0.640\\
\!\!\!\!0.279&\!\!\!\!0.592&\!\!\!\!0.756\\
\end{array}\right)$\!\!  \\\hline
$\sin^2 \theta_{12}$   &$\frac{2}{87}(28-5\sqrt{7})=0.3396$ & 
$ 0.321$&$0.341$ & $0.319$   
\\\hline
$\sin^2 \theta_{23}$   & $\frac{1}{29}(17-2\sqrt{7})=0.4037$&$0.389$& $0.376$ &  $0.417$
  \\\hline
$\sin^2 \theta_{13}$   &$\frac{1}{3}-\frac{5}{6\sqrt{7}}=0.01836$&$0.0215$ & $0.0211$ &$0.0186$    \\\hline
$m_1$  & $ (3-\sqrt{7})m$&$0.318m$& $0.330m$ &  $0.341m$  \\\hline
$m_2$  &$-m$ &$-1.03m$& $-0.977m$ & $-1.06m$   \\\hline
$m_3$   & $(3+\sqrt{7})m$& $5.28m$&$5.38m$ & $5.55m$   \\\hline
$\Delta m_{21}^2$   & $(-15+6\sqrt{7})m^2$& $0.966m^2$&$0.846m^2$ &  $1.00 m^2$  \\\hline
$\Delta m_{31}^2$   & $12\sqrt{7}m^2$&$27.7m^2$& $28.8m^2$ &  $30.6m^2$  \\\hline
\!\!\!\!\!\!$\begin{array}{c}|m|\vspace{-2mm}\\
( 10^{-3}{\footnotesize \eV})\\
\end{array}$\! \!\!\!\!\!\!\!&$ [8.95,~9.21]$ &$[9.06,~9.15]$&$[9.10,~9.67]$&
$[8.62,~8.97]$\\\hline
\end{tabular}%
} \caption{Examples of one-parameter neutrino mass matrix $M_{\nu}$ and 
related quantities for $|(M_{\nu})_{23}/(M_{\nu})_{13}|\simeq 3$ for the case of
 A$_1$. }
\label{tab1}
\end{table}

\begin{table}[h]
\centering {\fontsize{9}{12}
\begin{tabular}{|c||c|c|c|c|}
\hline  $M_{\nu}$ & 
$m\left(\begin{array}{ccc} 
0&0&1\\
0&\sqrt{6}&2\\
1&2&5/2\\
\end{array}\right)$ &
 $m\left(\begin{array}{ccc} 
0&0&1\\
0&5/2&2\\
1&2&5/2\\
\end{array}\right)$ &
 $m\left(\begin{array}{ccc} 
0&0&1\\
0&2\sqrt{2}&2\\
1&2&\sqrt{5}\\
\end{array}\right)$ &
 $m\left(\begin{array}{ccc} 
0&0&1\\
0&2\sqrt{2}&\sqrt{5}\\
1&\sqrt{5}&2\sqrt{2}\\
\end{array}\right)$  \\\hline
$U_{PMNS}$ &  
\!\!$\left(\begin{array}{ccc} 
\!\!\!\!0.825&\!\!\!\!0.543&\!\!\!\!0.157\\
\!\!\!\!0.312&\!\!\!\!-0.669&\!\!\!\!0.675\\
\!\!\!\!-0.472&\!\!\!\!0.507&\!\!\!\!0.721\\
\end{array}\right)$\!\! &
\!\!$\left(\begin{array}{ccc} 
\!\!\!\!0.828&\!\!\!\!0.539&\!\!\!\!0.156\\
\!\!\!\!0.306&\!\!\!\!-0.667&\!\!\!\!0.679\\
\!\!\!\!-0.470&\!\!\!\!0.515&\!\!\!\!0.717\\
\end{array}\right)$\!\!& 
\!\!$\left(\begin{array}{ccc} 
\!\!\!\!0.822&\!\!\!\!0.552&\!\!\!\!0.143\\
\!\!\!\!0.287&\!\!\!\!-0.618&\!\!\!\!0.732\\
\!\!\!\!-0.492&\!\!\!\!0.560&\!\!\!\!0.666\\
\end{array}\right)$ \!\! & 
\!\!$\left(\begin{array}{ccc} 
\!\!\!\!0.842&\!\!\!\!0.521&\!\!\!\!0.139\\
\!\!\!\!0.299&\!\!\!\!-0.664&\!\!\!\!0.685\\
\!\!\!\!-0.449&\!\!\!\!0.535&\!\!\!\!0.715\\
\end{array}\right)$\!\!  \\\hline
$\sin^2 \theta_{12}$   &$0.303$ & 
$ 0.297$&$0.311$ & $0.277$   
\\\hline
$\sin^2 \theta_{23}$   & $0.467$&$0.473$& $0.547$ &  $0.478$
  \\\hline
$\sin^2 \theta_{13}$   &$0.0247$&$0.0242$ & $0.0205$ &$0.0192$    \\\hline
$m_1$  & $ -0.572m$&$-0.567m$& $-0.599m$ &  $-0.533m$  \\\hline
$m_2$  &$0.933m$ &$0.956m$& $1.02m$ & $1.03m$   \\\hline
$m_3$   & $4.59m$& $4.61m$&$4.65m$ & $5.16m$   \\\hline
$\Delta m_{21}^2$   & $0.544m^2$& $0.592m^2$&$0.671m^2$ &  $0.70 m^2$  \\\hline
$\Delta m_{31}^2$   & $20.7m^2$&$20.9m^2$& $21.3m^2$ &  $26.4m^2$  \\\hline
\!\!\!\!\!\!$\begin{array}{c}|m|\vspace{-2mm}\\
( 10^{-3}{\footnotesize \eV})\\
\end{array}$\! \!\!\!\!\!\!\! &$ [11.3,~11.4]$ &$[10.9,~11.3]$&$[10.3,~11.0]$&
$[9.54,~10.1]$\\\hline
\end{tabular}%
} \caption{Examples of one-parameter neutrino mass matrix $M_{\nu}$ and 
related quantities for $|(M_{\nu})_{23}/(M_{\nu})_{13}|\simeq 2$ for the case of
 A$_1$. }
\label{tab1-1}
\end{table}
From the numerical calculation above, 
we find some examples of $M_{\nu}$ with one real free parameter $m$ 
as listed in Table \ref{tab1} for 
$|(M_{\nu})_{23}/(M_{\nu})_{13}|\simeq 3$ and in Table \ref{tab1-1} 
for $|(M_{\nu})_{23}/(M_{\nu})_{13}|\simeq 2$ for the case A$_1$
\footnote{See Ref.\cite{Araki:2012ip} for another example of 
the one-parameter texture 
$M_{\nu}=m\left( \begin{array}{ccc}
0&0&1\\0&3&2\\1&2&2\\
\end{array}\right)$.}.  
The bound of $|m|$ is obtained by the overlap of two constraints 
of $\Delta m_{21}^2$ and $\Delta m_{31}^2$ given in Eq.(\ref{exp}). 
One can see that all textures are in agreement with the current experimental bounds 
at the 3 $\sigma$ level given in Eq.(\ref{exp}). 
In addition to the mass matrices shown in Tables \ref{tab1} and \ref{tab1-1}, 
matrices with different sign are also candidates of one-parameter $M_{\nu}$, such 
as 
\bea
\left( \begin{array}{ccc}
0&0&1\\
0&2&3\\
1&3&3\\
\end{array}\right)\to
\left( \begin{array}{ccc}
0&0&1\\
0&2&-3\\
1&-3&3\\
\end{array}\right),~
\left( \begin{array}{ccc}
0&0&1\\
0&-2&3\\
1&3&-3\\
\end{array}\right),~
\left( \begin{array}{ccc}
0&0&1\\
0&-2&-3\\
1&-3&-3\\
\end{array}\right),
\label{sign}
\eea
with the same observables $\sin^2 \theta_{12,23,13}$, $\Delta m^2_{21,31}$ 
and consequently the same bounds of $|m|$, while the sign of mass eigenvalues $m_{1,2,3}$ and that of 
the PMNS matrix 
$U_{PMNS}$ are different.  
For the case A$_2$, all mass matrices $M_{\nu}({\rm A}_2)$ obtained from 
$M_{\nu}({\rm A}_1)$ listed in Tables \ref{tab1}, \ref{tab1-1} and Eq.(\ref{mnu0}), 
are also consistent with Eq.(\ref{exp}). The PMNS matrix for the case A$_2$ is 
given from that for the case A$_1$ by 
$U_{PMNS}({\rm A}_2)=PU_{PMNS}({\rm A}_1)$, {\it i.e.},
$\sin^2 \theta_{23}({\rm A}_2)=1-\sin^2 \theta_{23}({\rm A}_1) $ and the other 
quantities remain unchanged.

In the next section, we discuss symmetry realization of the mass matrix 
\be
M_{\nu}=m\left( \begin{array}{ccc}
0&0&1\\
0&2&3\\
1&3&3\\
 \end{array}\right),
 \label{mnu}
\ee
by the binary icosahedral symmetry $A_5'$, as an example. 

\section{Symmetry Realization}
In this section, we discuss symmetry realization of the one-parameter texture 
Eq.(\ref{mnu}) by the binary icosahedral symmetry $A_5'$ 
and the Higgs potential of our model.
\subsection{Mass Matrices}

\begin{table}[thbp]
\centering {\fontsize{10}{12}
\begin{tabular}{||c|c|c||c|c|c|c|c||}
\hline\hline ~~~~ & ~~ $L_a$~~ & ~~$e^c_a$~~ & ~~ $\Phi_0$~~
  & $\Phi_a$  & $\Phi'_A$ &$\Delta_A$ \\\hline
$(SU(2)_L,U(1)_Y)$ &  $({\bf{2}},-1/2)$ & $({\bf{1}},1)$  & $({\bf{2}},1/2)$  & $({\bf{2}},
1/2)$ & $({\bf{2}},1/2)$ & $({\bf{3}},1)$  \\\hline
$A_5'$   & ${\bf 3}_a$ & ${\bf 3}_a$ & ${\bf 1}$  & ${\bf 3}_a$ & ${\bf 5}_A$& ${\bf 5}
_A$\\\hline
\end{tabular}%
} \caption{The particle contents ($a=1-3$ and $A=1-5$). }
\label{tab2}
\end{table}

Now we discuss symmetry realization of the texture given in 
Eq.(\ref{mnu}) for neutrino mass matrix. We consider 
the binary icosahedral symmetry $A_5'$ \cite {Hashimoto:2011tn} as 
a flavor symmetry. The $A_5'$ symmetry contains three- and five-dimensional 
irreducible representations ${\bf 3}$ and ${\bf 5}$, and their tensor product 
gives $A_5'$ invariance ${\bf 1}$. Moreover the resultant singlet ${\bf 1}$ from 
${\bf 3 \cdot 5 \cdot 3}$ enters all the elements of $M_{\nu}$ with 
desired weights if 
neutrinos and Higgs bosons are embedded into ${\bf 3}$ and ${\bf 5}$, respectively. 
Therefore the $A_5'$ symmetry is preferable for the one-parameter texture.   
 For the lepton sector, the particle contents are  
shown in Table \ref{tab2}, where $L_a,~e^c_a,~(\Phi_{0,a},~\Phi'_A)$ and $\Delta_A$ ($a=1-3,~A=1-5$) are the $SU(2)_L$ doublet leptons, 
$SU(2)$ singlet leptons, $SU(2)$ doublet Higgs fields and $SU(2)$ triplet 
Higgs fields, respectively. Since right-handed neutrinos are absent in 
our model, the $SU(2)_L$ doublet Higgs fields are responsible only for masses of 
charged leptons, while neutrino masses are generated by the 
vacuum expectation values (VEVs) of $SU(2)_L$ 
triplet Higgs fields through the type-II seesaw mechanism. If one assigns 
${\bf 1}$ for quarks, only $\Phi_0$ couples to quarks and gives their masses. 
Although there are no predictions in the quark sector, 
it is ensured to be the same as the standard model.  

Now we give the multiplication rules of the $A_5'$ group relevant for the Yukawa 
interactions in 
our model. For ${\bf 3}$ and ${\bf 5}$ irreducible representations
\bea
&&{\bf 3}=\left( \begin{array}{c} 
x_1\\x_2\\x_3\\
\end{array}\right),~
\left( \begin{array}{c} 
y_1\\y_2\\y_3\\
\end{array}\right),~
{\bf 5}=\left( \begin{array}{c} 
X_1\\X_2\\X_3\\X_4\\X_5\\
\end{array}\right),~
\left( \begin{array}{c} 
Y_1\\Y_2\\Y_3\\Y_4\\Y_5\\
\end{array}\right),
\eea
their tensor products 
are given by \footnote{See Ref.\cite{Hashimoto:2011tn} for the complete multiplication 
rules.}
\bea
&&
{\bf 3}\otimes {\bf 3}
=\left( x_1 y_3-x_2 y_2+x_3 y_1\right)_{{\bf 1}}
+\frac{1}{\sqrt{2}}\left( \begin{array}{c}
x_1 y_2-x_2y_1\\
x_1y_3-x_3y_1\\
x_2y_3-x_3y_2
\end{array}\right)_{{\bf 3}}
+\left(\begin{array}{c} 
x_1 y_1\\
\frac{1}{\sqrt{2}}\left(x_1y_2+x_2y_1 \right)\\
\frac{1}{\sqrt{6}}\left( x_1y_3+2x_2y_2+x_3y_1\right)\\
\frac{1}{\sqrt{2}}\left( x_2y_3+x_3y_2\right)\\
x_3y_3\\
\end{array}\right)_{{\bf 5}},\nn\\&&
{\bf 5}\otimes {\bf 5}
=\left( X_1Y_5-X_2Y_4+X_3Y_3-X_4Y_2+X_5Y_1\right)_{\bf 1}+\cdots.
\label{rule}
\eea

From the particle contents shown in Table \ref{tab2} and the 
multiplication rules Eq. (\ref{rule}), the $A_5'$ invariant Yukawa interactions are
 given by
\bea
{\cal L}_Y&=&y_{\Delta}L_a^T i \sigma_2 \Delta_A L_b
+y_1 \Phi_0^{\dag}L_a e_b^c
+y_2  \Phi_a^{\dag}L_b e_c^c 
+y_3 {\Phi'}_A^{\dag}L_a e_b^c+c.c.,
\label{yukawa}
\eea
where all indices are summed up in $A_5'$ invariant way in 
accordance with Eq(\ref{rule}). 
After the electroweak symmetry breaking by the VEVs of the Higgs fields defined by
\be
\langle \Delta_A \rangle=\frac{1}{\sqrt{2}}v_{\Delta A},~~
\langle \Phi_{0,a} \rangle=\frac{1}{\sqrt{2}}v_{0,a},~~
\langle \Phi'_{A} \rangle=\frac{1}{\sqrt{2}}V_{A},
\ee
with $v_{0}^2+\sum_a v_a^2+\sum_A(V_A^2+2v_{\Delta A}^2)=(246\GeV)^2$
, we obtain the following mass matrix
\be
M_{\nu}=\frac{y_{\Delta}}{\sqrt{2}}\left( \begin{array}{ccc}
v_{\Delta 5}&-\frac{1}{\sqrt{2}}v_{\Delta 4}&\frac{1}{\sqrt{6}}v_{\Delta 3}\\
-\frac{1}{\sqrt{2}}v_{\Delta 4}&\frac{2}{\sqrt{6}}v_{\Delta 3}&
-\frac{1}{\sqrt{2}}v_{\Delta 2}\\
\frac{1}{\sqrt{6}}v_{\Delta 3}&-\frac{1}{\sqrt{2}}v_{\Delta 2}&
v_{\Delta 1}\\
\end{array}\right),
\label{mnu2}
\ee
for neutrino sector, and
\be
M_e=\frac{1}{\sqrt{2}}\left( \begin{array}{ccc}
y_3 V_5&\frac{1}{\sqrt{2}}y_2 v_3-\frac{1}{\sqrt{2}}y_3V_4&y_1 v_0
-\frac{1}{\sqrt{2}}y_2v_2+\frac{1}{\sqrt{6}}y_3 V_3\\
-\frac{1}{\sqrt{2}}y_2 v_3-\frac{1}{\sqrt{2}}y_3V_4&
-y_1 v_0+\frac{2}{{\sqrt{6}}}y_3V_3&\frac{1}{\sqrt{2}}y_2 v_1-\frac{1}{\sqrt{2}}y_3 V_2\\
y_1v_0+\frac{1}{\sqrt{2}}y_2 v_2+\frac{1}{\sqrt{6}}y_3V_3&
-\frac{1}{\sqrt{2}}y_2 v_1-\frac{1}{\sqrt{2}}y_3 V_2&y_3 V_1\\
\end{array}\right),
\label{me2}
\ee
for charged lepton sector. 
If the Higgs fields obtain the following VEVs, 
\bea
&&v_{\Delta 1}=3 v_{\Delta},~v_{\Delta 2}=-3\sqrt{2} v_{\Delta},~
v_{\Delta 3}=\sqrt{6} v_{\Delta},~v_{\Delta 4}=v_{\Delta 5}=0,\nn\\&&
v_0\neq 0,~v_2\neq 0,~V_3\neq 0,~v_1=v_3=V_1=V_2=V_4=V_5=0,~
\label{vev}
\eea
one finds that the mass matrices $M_{\nu}$ and $M_e$ have the desired form
\be
M_{\nu}=\frac{y_{\Delta}v_{\Delta}}{\sqrt{2}}
\left( \begin{array}{ccc}
0&0&1\\
0&2&3\\
1&3&3\\
\end{array}\right),~~~
M_e=\left(\begin{array}{ccc}
0&0&m_e\\
0&-m_{\mu}&0\\
m_{\tau}&0&0\\
\end{array} \right),
\label{mnume}
\ee
where
\bea
&&
m_e=\frac{1}{\sqrt{2}}\left( y_1 v_0-\frac{1}{\sqrt{2}}y_2v_2+\frac{1}{\sqrt{6}}y_3V_3\right),\\&&
m_{\mu}=\frac{1}{\sqrt{2}}\left(y_1v_0-\frac{2}{\sqrt{6}}y_3V_3 \right),\\&&
m_{\tau}=\frac{1}{\sqrt{2}}\left( y_1 v_0+\frac{1}{\sqrt{2}}y_2v_2+\frac{1}{\sqrt{6}}y_3V_3\right), 
\eea
and $M_eM_e^{\dag}={\rm diag}(m_e^2,~m_{\mu}^2,~m_{\tau}^2)$. The VEVs of the triplet Higgs $\Delta_A$ gives 
additional contributions to the $\rho$ parameter 
\be
\rho=\frac{v^2}
{v^2+2\sum_{A} v_{\Delta A}^2}
= \frac{v^2}{v^2+66 v_{\Delta}^2}, 
\ee
where $v^2=v_0^2+\sum_a v_a^2+\sum_A V_A^2=v_0^2+v_2^2+V_3^2$ 
because of Eq. (\ref{vev}). The experimental value 
$\rho_{\rm exp}=1.0004^{+0.0003}_{-0.0004}$ constrains $v_{\Delta}$ to be 
smaller than about $0.61\GeV$ at the 95$\%$ confidence level. 
Therefore we assume $v_{\Delta}\ll v_0,v_2,V_3$. 
As already discussed in the last section, the mass matrices Eq.(\ref{mnume}) 
with $m=y_{\Delta}v_{\Delta}/\sqrt{2}$ corresponding to Eq.(\ref{mnu})  are 
compatible with the current experimental bounds.

\subsection{Higgs Sector}
As for the Higgs sector, the total Higgs potential is given in Appendix in symbolic form. Here we mention the tadpole conditions. The $A'_5$ invariant scalar mass terms are 
given by 
\bea
&&V_{\rm mass}=m_0^2 \Phi_0^{\dag}\Phi_0
+m_3^2\left( \Phi_1^{\dag}\Phi_3-\Phi_2^{\dag}\Phi_2+\Phi_3^{\dag}\Phi_1 \right)
\nn\\&&
+m_5^2 \left( {\Phi'}_1^{\dag}\Phi'_5- {\Phi'}_2^{\dag}\Phi'_4+
 {\Phi'}_3^{\dag}\Phi'_3- {\Phi'}_4^{\dag}\Phi'_2+ {\Phi'}_5^{\dag}\Phi'_1\right)\nn\\&&
 +m_{\Delta}^2{\rm Tr}\left[ \Delta_1^{\dag}\Delta_5- \Delta_2^{\dag}\Delta_4
 + \Delta_3^{\dag}\Delta_3- \Delta_4^{\dag}\Delta_2+ \Delta_5^{\dag}\Delta_1\right]. 
 \label{scalarmass}
\eea
Since the total $A'_5$ invariant potential does not have any accidental symmetry, 
our model does not suffer from the problem of massless Goldstone bosons. 

By imposing the conditions for the VEVs Eq.(\ref{vev}), the tadpole conditions give the scalar masses 
\bea
m_0^2&\simeq&-\lambda_1 v_0^2+\left( \frac 12 \tilde \epsilon_1
-\frac{V_3}{\sqrt{6}v_0}\tilde \kappa_8\right)v_2^2
-\frac 12\left( \tilde \epsilon_2-\frac{V_3}{v_0}\kappa_2^{(1)}\right)V_3^2,
\label{higgsm1}\\
m_{3}^2 &\simeq&
-\frac 12 \tilde \epsilon_1 v_0^2
+\left( \lambda_2^{(1)}+\frac 23 \lambda_2^{(3)}\right)v_2^2
-\left( \tilde \epsilon_4+2 \tilde \epsilon_5 +\tilde \epsilon_6\right)V_3^2
+\frac{\sqrt{6}}{3}\tilde \kappa_8 v_0V_3
-2\sqrt{2}v_{\Delta}\mu_1,\label{higgsm2}\\
m_{5}^2&\simeq&
-\frac 12 \tilde \epsilon_2 v_0^2
+\left( \tilde \epsilon_4+2 \tilde \epsilon_5 +\tilde \epsilon_6
-\frac{v_0}{\sqrt{6}V_3}\tilde \kappa_8\right)v_2^2
-\tilde \lambda_3 V_3^2
+\frac 32 \kappa_2^{(1)}v_0 V_3
-2\sqrt{3}v_{\Delta}\left( \mu_2-6 \mu_3\right),\label{higgsm3}\\
m_{\Delta}^2&\simeq&
-\frac 12 \tilde \epsilon_3 v_0^2
+\left( \tilde \epsilon_{7}+ \tilde \epsilon_{8}+\frac{\mu_1}{3\sqrt{2}v_{\Delta}}\right)v_2^2
-\frac 12\left( \tilde \lambda_4+\tilde \lambda_5
+\frac{\mu_2-6\mu_3}{\sqrt{3}v_{\Delta}}\right)V_3^2
+\tilde \kappa_2 v_0 V_3,
\label{higgsm4}
\eea
where 
\bea
\tilde \lambda_i&=&\lambda_i^{(1)}-\lambda_i^{(7)}-6\lambda_i^{(8)}+36\lambda_i^{(9)}~
(i=3,4,5),\nn\\
\tilde \epsilon_{i}&=&\epsilon_{i}^{(1)}+\epsilon_{i}^{(2)}
+2 \epsilon_{i}^{(3)}~(i=1,2),~
\tilde \epsilon_3=\epsilon_3^{(1)}+\epsilon_3^{(2)},~
\tilde \epsilon_{i}=\frac 12 \epsilon_i^{(1)}+\frac{1}{\sqrt{6}}\epsilon_i^{(3)}
+\frac{3}{10}\epsilon_i^{(4)}~(i=4,5,6,7,8),\nn\\
\tilde \kappa_2&=&\kappa_2^{(2)}+\kappa_2^{(3)},~
\tilde \kappa_8=\kappa_8^{(1)}+\kappa_8^{(2)}+\kappa_8^{(3)}. 
\label{lambda}
\eea
The coupling constants $\lambda$s, $\epsilon$s and $\kappa$s are 
given in Appendix. 
In Eqs. (\ref{higgsm1})-(\ref{higgsm4}), the terms proportional to $v_{\Delta}^2$ 
have been neglected. 
The pattern 
\be
M_{\nu}=\left( \begin{array}{ccc}
0&0&1\\
0&2&3\\
1&3&2\sqrt{2}\\
\end{array}\right),
\ee
listed in Table \ref{tab1}  
can be realized by the $A_5'$ symmetry in similar fashion, with 
corresponding tadpole conditions. 
\section{Conclusions}
We have studied neutrino mass matrix $M_{\nu}$ with two texture zeros and 
its symmetry realization. After confirming that the cases A$_{1}$ and A$_2$ 
satisfy the current experimental constraints by numerical calculation, we 
have found some examples of $M_{\nu}$ with {\it one} real parameter. 
Since the magnitude of non-zero elements of $M_{\nu}$ 
is restricted in small region, one can extract examples of $M_{\nu}$ with 
simple forms. While there exist infinite number of candidates of one-parameter 
$M_{\nu}$, such simple forms are preferable in the standpoint of flavor symmetry. 

Next we have discussed symmetry realization of one-parameter $M_{\nu}$ 
and the Higgs potential based on the 
binary icosahedral symmetry $A_5'$. 
The $A_5'$ symmetry contains three- and five-dimensional 
irreducible representations, and their tensor product ${\bf 3 \cdot 5 \cdot 3}$ 
enters all the elements of $M_{\nu}$ with definite weights. 
If one assigns ${\bf 5}$ to $SU(2)_L$ triplet Higgs $\Delta$, 
desired neutrino mass matrix is obtained by the type-II seesaw mechanism and 
by choosing the vacua of the Higgs potential. 
While we have shown one example of symmetry realization in this paper, 
the $A_5'$ symmetry can work for the other one-parameter $M_{\nu}$ because of 
its multiplication rules.

\section*{Acknowledgments}
We would like to thank H. Okada and A. Shibuya for useful discussions.

\begin{appendix}
\section{Higgs Potential}
Here we show the Higgs potential of our model with the A$_5'$
symmetry. In order to avoid redundancy, we give a symbolic form of the potential.
The Higgs fields are denoted by
\be
\Phi_0\equiv {\bf 1},~\Phi_a\equiv {\bf 3},~\Phi'_A\equiv {\bf 5},~
\Delta_A \equiv \Delta
\ee
where $a=1-3$ and $A=1-5$ for three- and five- 
dimensional representation, respectively. Their Hermitian conjugate fields are 
denoted by $\Phi_a^{\dag} \equiv {\bf 3}^{\dag}$ etc., while 
${\bf 3}^{\dag}$ and ${\bf 3}$ obey the same multiplication rules. 
We represent a product
\bea
{\bf 3}\otimes {\bf 3} ={\bf 1}\oplus {\bf 3} \oplus {\bf 5}~&\rightarrow&~
{\bf 33}=({\bf 33})_{\bf 1}+({\bf 33})_{\bf 3}+({\bf 33})_{\bf 5}
=({\bf 33})_{\alpha},\nn\\
&{\rm or}&~{\bf 3^{\dag}3}=({\bf 3^{\dag}3})_{\bf 1}+({\bf 3^{\dag}3})_{\bf 3}
+({\bf 3^{\dag}3})_{\bf 5}=({\bf 3^{\dag}3})_{\alpha},
\eea
depending on the gauge quantum number of the Higgs fields. 
The index $\alpha$ must be summed up for all possible combinations of 
the tensor product. For example in the case of $({\bf 3^{\dag} 3})^2$, since only the products ${\bf  11},~{\bf 33}$ and ${\bf 55}$ 
can be invariant under A$_5'$, we 
denote 
$\lambda_2^{(\alpha)}({\bf 3}^{\dag}{\bf 3})_{\alpha}({\bf 3}^{\dag}{\bf 3})_{ \alpha}$ 
as 
\bea
&&({\bf 3^{\dag} 3})^2~:~
\lambda_2^{(\alpha)}({\bf 3}^{\dag}{\bf 3})_{\alpha}({\bf 3}^{\dag}{\bf 3})_{ \alpha}
=\lambda_2^{(1)}({\bf 3}^{\dag}{\bf 3})_{\bf 1}({\bf 3}^{\dag}{\bf 3})_{ \bf 1}
+\lambda_2^{(2)}({\bf 3}^{\dag}{\bf 3})_{\bf 3}({\bf 3}^{\dag}{\bf 3})_{ \bf 3}
+\lambda_2^{(3)}({\bf 3}^{\dag}{\bf 3})_{\bf 5}({\bf 3}^{\dag}{\bf 3})_{ \bf 5}\nn\\&&
=\lambda_2^{(1)}\left( \Phi_1^{\dag}\Phi_3-\Phi_2^{\dag}\Phi_2+\Phi_3^{\dag}\Phi_1\right)^2\nn\\&&
+\lambda_2^{(2)}\left[2\left(\Phi_1^{\dag}\Phi_2-\Phi_2^{\dag}\Phi_1 \right) 
\left(\Phi_2^{\dag}\Phi_3-\Phi_3^{\dag}\Phi_2 \right)
-\left(\Phi_1^{\dag}\Phi_3-\Phi_3^{\dag}\Phi_1 \right)^2 \right]\nn\\&&
+\lambda_2^{(3)}\left[ 
2\left(\Phi_1^{\dag}\Phi_1\right)\left(\Phi_3^{\dag}\Phi_3\right)
-\left(\Phi_1^{\dag}\Phi_2+\Phi_2^{\dag}\Phi_1 \right) 
\left(\Phi_2^{\dag}\Phi_3+\Phi_3^{\dag}\Phi_2 \right)
+\frac{1}{6}\left( \Phi_1^{\dag}\Phi_3+\Phi_2^{\dag}\Phi_2+\Phi_3^{\dag}\Phi_1\right)^2
\right].\nn\\
&&
\eea
Moreover in what follows, trace for $\Delta$ is omitted, and ``~$\cdot$~'' denotes the Pauli matrices 
$\sigma^{1,2,3}$. 

In the notation described above, the Higgs potential except the bilinear terms 
given in the main text can be written down as 
\bea
{\bf 3}^2{\bf 5},{\bf 5}^3,{\bf 3}{\bf 5}^2&:&
\mu_1{\bf 3} \Delta^{\dag} {\bf 3}
+\mu_2({\bf 5} \Delta^{\dag} )_{\bf 5_1}{\bf 5}
+\mu_3({\bf 5} \Delta^{\dag})_{\bf 5_2} {\bf 5}
+\mu_4{\bf 3} \Delta^{\dag} {\bf 5}+h.c.,
\eea
for the trilinear terms, and 
\bea
{\bf 1}^4&:&\lambda_1 ({\bf 1}^{\dag} {\bf 1})^2,\nn \\
{\bf 3}^4&:&\lambda_2^{(\alpha)}({\bf 3}^{\dag}{\bf 3})_{\alpha}
({\bf 3}^{\dag}{\bf 3})_{ \alpha},\nn\\
{\bf 5}^4&:&
\lambda_3^{(\alpha)}({\bf 5}^{\dag}{\bf 5})_{\alpha}({\bf 5}^{\dag}{\bf 5})_{\alpha}
+\lambda_4^{(\alpha)}({\bf 5}^{\dag}{\bf 5})_{\alpha}({\Delta}^{\dag}{\Delta})_{\alpha}
+\lambda_5^{(\alpha)}({\bf 5}^{\dag}\cdot{\bf 5})_{\alpha}({\Delta}^{\dag}
\cdot{\Delta})_{\alpha}\nn\\
&&+\lambda_6^{(\alpha)}((\Delta^{\dag}\Delta)_{\alpha})^2
+\lambda_7\det\left[ \Delta^{\dag}\Delta\right]
,\nn\\
{\bf 1}^2{\bf 3}^2&:&
\epsilon_1^{(1)}({\bf 1}^{\dag}{\bf 1})_{{\bf 1}}({\bf 3}^{\dag}{\bf 3})_{{\bf 1}}
+\epsilon_1^{(2)}({\bf 1}^{\dag}{\bf 3})_{{\bf 3}}({\bf 3}^{\dag}{\bf 1})_{{\bf 3}}
+\epsilon_1^{(3)}\left[( ({\bf 1}^{\dag}{\bf 3})_{{\bf 3}})^2+h.c.\right],\nn\\
{\bf 1}^2{\bf 5}^2&:&
\epsilon_2^{(1)}({\bf 1}^{\dag}{\bf 1})_{{\bf 1}}({\bf 5}^{\dag}{\bf 5})_{{\bf 1}}
+\epsilon_2^{(2)}({\bf 1}^{\dag}{\bf 5})_{{\bf 5}}({\bf 5}^{\dag}{\bf 1})_{{\bf 5}}
+\epsilon_2^{(3)}\left[( ({\bf 1}^{\dag}{\bf 5})_{{\bf 5}})^2+h.c.\right]\nn\\
&&+\epsilon_3^{(1)}({\bf 1}^{\dag}{\bf 1})_{{\bf 1}}(\Delta^{\dag}\Delta)_{\bf 1}
+\epsilon_3^{(2)}({\bf 1}^{\dag}\cdot{\bf 1})_{{\bf 1}}(\Delta^{\dag}\cdot\Delta)_{\bf 1},
\nn\\
{\bf 3}^2{\bf 5}^2&:&
\epsilon_4^{(\alpha)}({\bf 3}^{\dag}{\bf 3})_{\alpha}({\bf 5}^{\dag}{\bf 5})_{\alpha}
+\epsilon_5^{(\alpha)}\left[(({\bf 3}^{\dag}{\bf 5})_{\alpha})^2+h.c.\right]
+\epsilon_6^{(\alpha)}({\bf 3}^{\dag}{\bf 5})_{\alpha}({\bf 5}^{\dag}{\bf 3})_{\alpha}\nn\\
&&+\epsilon_{7}^{(\alpha)}({\bf 3}^{\dag}{\bf 3})_{\alpha}(\Delta^{\dag}\Delta)_{\alpha}
+\epsilon_{8}^{(\alpha)}({\bf 3}^{\dag}\cdot{\bf 3})_{\alpha}(\Delta^{\dag}\cdot
\Delta)_{\alpha}\nn\\
{\bf 1}{\bf 3}^3&:&\kappa_1 ({\bf 1}^{\dag}{\bf 3})_{\bf 3}({\bf 3}^{\dag}{\bf 3})_{\bf 3}
+h.c.,\nn\\
{\bf 1}{\bf 5}^3&:&
\kappa_2^{(1)}({\bf 1}^{\dag}{\bf 5})_{\bf 5}({\bf 5}^{\dag}{\bf 5})_{\bf 5}
+\kappa_2^{(2)}({\bf 1}^{\dag}{\bf 5})_{\bf 5}(\Delta^{\dag}\Delta)_{\bf 5}
+\kappa_2^{(3)}({\bf 1}^{\dag}\cdot{\bf 5})_{\bf 5}(\Delta^{\dag}\cdot\Delta)_{\bf 5}
+h.c.,\nn\\
{\bf 3}^3{\bf 5}&:&
\kappa_3^{(1)}({\bf 3}^{\dag}{\bf 3})_{\bf 3}({\bf 3}^{\dag}{\bf 5})_{\bf 3}
+\kappa_3^{(2)}({\bf 3}^{\dag}{\bf 3})_{\bf 5}({\bf 3}^{\dag}{\bf 5})_{\bf 5}+h.c.,\nn\\
{\bf 3}{\bf 5}^3&:&
\kappa_4^{(\alpha)}({\bf 3}^{\dag}{\bf 5})_{\alpha}({\bf 5}^{\dag}{\bf 5})_{\alpha}
+\kappa_5^{(\alpha)}({\bf 3}^{\dag}{\bf 5})_{\alpha}(\Delta^{\dag}\Delta)_{\alpha}
+\kappa_6^{(\alpha)}({\bf 3}^{\dag}\cdot{\bf 5})_{\alpha}(\Delta^{\dag}\cdot\Delta)_{\alpha}+h.c.,\nn\\
{\bf 1}{\bf 3}{\bf 5}^2&:&
\kappa_7^{(1)}({\bf 1}^{\dag}{\bf 3})_{\bf 3}({\bf 5}^{\dag}{\bf 5})_{\bf 3}
+\kappa_7^{(2)}({\bf 1}^{\dag}{\bf 5})_{\bf 5}({\bf 3}^{\dag}{\bf 5})_{\bf 5}
+\kappa_7^{(3)}({\bf 1}^{\dag}{\bf 5})_{\bf 5}({\bf 5}^{\dag}{\bf 3})_{\bf 5}\nn\\
&&+\kappa_7^{(4)}({\bf 1}^{\dag}{\bf 3})_{\bf 3}(\Delta^{\dag}\Delta)_{\bf 3}
+\kappa_7^{(5)}({\bf 1}^{\dag}\cdot{\bf 3})_{\bf 3}(\Delta^{\dag}\cdot\Delta)_{\bf 3}
+h.c.,\nn\\
{\bf 1}{\bf 3}^2{\bf 5}&:&
\kappa_{8}^{(1)}({\bf 1}^{\dag}{\bf 3})_{\bf 3}({\bf 3}^{\dag}{\bf 5})_{\bf 3}
+\kappa_{8}^{(2)}({\bf 1}^{\dag}{\bf 3})_{\bf 3}({\bf 5}^{\dag}{\bf 3})_{\bf 3}
+\kappa_{8}^{(3)}({\bf 1}^{\dag}{\bf 5})_{\bf 5}({\bf 3}^{\dag}{\bf 3})_{\bf 5}
+h.c.,
\eea
for the quartic terms. In the above expressions, the sums of index $\alpha$ are
 defined as 
\bea
&&{\bf 3}^4~:~
\lambda_2^{(\alpha)}({\bf 3}^{\dag}{\bf 3})_{\alpha}({\bf 3}^{\dag}{\bf 3})_{ \alpha}
=\lambda_2^{(1)}({\bf 3}^{\dag}{\bf 3})_{\bf 1}({\bf 3}^{\dag}{\bf 3})_{ \bf 1}
+\lambda_2^{(2)}({\bf 3}^{\dag}{\bf 3})_{\bf 3}({\bf 3}^{\dag}{\bf 3})_{ \bf 3}
+\lambda_2^{(3)}({\bf 3}^{\dag}{\bf 3})_{\bf 5}({\bf 3}^{\dag}{\bf 3})_{ \bf 5}\nn\\&&
{\bf 5}^4~:~
\lambda^{(\alpha)}({\bf 5}^{\dag}{\bf 5})_{\alpha}({\bf 5}^{\dag}{\bf 5})_{ \alpha}
=\lambda^{(1)}({\bf 5}^{\dag}{\bf 5})_{\bf 1}({\bf 5}^{\dag}{\bf 5})_{\bf 1}
+\lambda^{(2)}({\bf 5}^{\dag}{\bf 5})_{\bf 3}({\bf 5}^{\dag}{\bf 5})_{\bf 3}
+\lambda^{(3)}({\bf 5}^{\dag}{\bf 5})_{\bf 3'}({\bf 5}^{\dag}{\bf 5})_{\bf 3'}\nn\\&&
\hspace{4.3cm}
+\lambda^{(4)}({\bf 5}^{\dag}{\bf 5})_{\bf 4_1}({\bf 5}^{\dag}{\bf 5})_{\bf 4_1}
+\lambda^{(5)}({\bf 5}^{\dag}{\bf 5})_{\bf 4_1}({\bf 5}^{\dag}{\bf 5})_{\bf 4_2}
+\lambda^{(6)}({\bf 5}^{\dag}{\bf 5})_{\bf 4_2}({\bf 5}^{\dag}{\bf 5})_{\bf 4_2}\nn\\&&
\hspace{4.3cm}
+\lambda^{(7)}({\bf 5}^{\dag}{\bf 5})_{\bf 5_1}({\bf 5}^{\dag}{\bf 5})_{\bf 5_1}
+\lambda^{(8)}({\bf 5}^{\dag}{\bf 5})_{\bf 5_1}({\bf 5}^{\dag}{\bf 5})_{\bf 5_2}
+\lambda^{(9)}({\bf 5}^{\dag}{\bf 5})_{\bf 5_2}({\bf 5}^{\dag}{\bf 5})_{\bf 5_2}\nn\\&&
{\bf 3}^2{\bf 5}^2~:~
\epsilon^{(\alpha)}({\bf 3^{\dag}3})_{\alpha}({\bf 5^{\dag}5})_{\alpha}
=\epsilon^{(1)}({\bf 3^{\dag}3})_{\bf 1}({\bf 5^{\dag}5})_{\bf 1}
+\epsilon^{(2)}({\bf 3^{\dag}3})_{\bf 3}({\bf 5^{\dag}5})_{\bf 3}
+\epsilon^{(3)}({\bf 3^{\dag}3})_{\bf 5}({\bf 5^{\dag}5})_{\bf 5}\nn\\&&
\hspace{4.6cm}
+\epsilon^{(4)}({\bf 3^{\dag}3})_{\bf 3'}({\bf 5^{\dag}5})_{\bf 3'}
+\epsilon^{(5)}({\bf 3^{\dag}3})_{\bf 4}({\bf 5^{\dag}5})_{\bf 4}\nn\\&&
{\bf 3}{\bf 5}^3~:~
\kappa^{(\alpha)}({\bf 3}^{\dag}{\bf 5})_{\alpha}({\bf 5}^{\dag}{\bf 5})_{\alpha}
=\kappa^{(1)}({\bf 3}^{\dag}{\bf 5})_{\bf 3}({\bf 5}^{\dag}{\bf 5})_{\bf 3}
+\kappa^{(2)}({\bf 3}^{\dag}{\bf 5})_{\bf 3'}({\bf 5}^{\dag}{\bf 5})_{\bf 3'}
+\kappa^{(3)}({\bf 3}^{\dag}{\bf 5})_{\bf 4}({\bf 5}^{\dag}{\bf 5})_{\bf 4_1}\nn\\&&
\hspace{4.6cm}
+\kappa^{(4)}({\bf 3}^{\dag}{\bf 5})_{\bf 4}({\bf 5}^{\dag}{\bf 5})_{\bf 4_2}
+\kappa^{(5)}({\bf 3}^{\dag}{\bf 5})_{\bf 5}({\bf 5}^{\dag}{\bf 5})_{\bf 5_1}
+\kappa^{(6)}({\bf 3}^{\dag}{\bf 5})_{\bf 5}({\bf 5}^{\dag}{\bf 5})_{\bf 5_2}.
\eea
Here notice that the product ${\bf 55}$ contains two ${\bf 4}(={\bf 4_1,4_2})$ and 
${\bf 5}(={\bf 5_1,5_2})$. See Ref.\cite{Hashimoto:2011tn} for details.
\end{appendix}

\end{document}